\documentclass[aps,pre,twocolumn,superscriptaddress,showkeys]{revtex4}
\usepackage{graphicx}

\begin{document}

\title[Fast ignition driven by quasi-monoenergetic ions]
{Fast ignition driven by quasi-monoenergetic ions: Optimal ion type
and reduction of ignition energies with an ion beam array}

\author{J.J.~Honrubia}
\email{javier.honrubia@upm.es}
\affiliation{ETSI Aeron\'auticos, Universidad Polit\'ecnica de Madrid, Madrid, Spain}

\author{J.C.~Fern\'andez}
\affiliation{Los Alamos National Laboratory, Los Alamos, New Mexico, USA}

\author{B.M.~Hegelich}
\affiliation{Los Alamos National Laboratory, Los Alamos, New Mexico, USA}

\author{M.~Murakami}
\affiliation{Institute of Laser Engineering, Osaka University, Osaka, Japan}

\author{C.D.~Enriquez}
\affiliation{ETSI Aeron\'auticos, Universidad Polit\'ecnica de Madrid, Madrid, Spain}

\date{\today}

\begin{abstract}
Fast ignition of inertial fusion targets driven by quasi-monoenergetic ion beams is
investigated by means of numerical simulations. Light and intermediate ions
such as lithium, carbon, aluminum and vanadium have been considered. Simulations
show that the minimum ignition energies of an ideal configuration of compressed
Deuterium-Tritium are almost independent on the ion atomic number. However,
they are obtained for increasing ion energies, which scale, approximately,
as $Z^2$, where $Z$ is the ion atomic number. Assuming that the ion beam
can be focused into 10 $\mu$m spots, a new irradiation scheme is proposed to
reduce the ignition energies. The combination of intermediate Z ions, such
as 5.5 GeV vanadium, and the new irradiation scheme allows a reduction of
the number of ions required for ignition by, roughly, three orders of
magnitude when compared with the standard proton fast ignition scheme.
\end{abstract}


\maketitle

\section{Introduction}

Fast ignition (FI) by laser-driven ion beams has been proposed
(Tabak and Callaham-Miller 1998, Roth et al. 2001 and Roth 2009)
as an alternative to the standard electron fast ignition scheme
(Tabak et al. 1994, Robinson et al. 2014).
This last scheme is limited by the high electron divergences
and kinetic energies observed in experiments (Green et al. 2008)
and PIC simulations (Debayle et al. 2010, Kemp and Divol 2012).
In addition, it is very sensitive to the energy level of the
laser ASE (Amplified Spontaneous Emission) pre-pulses
(Baton et al. 2008, Shiraga et al. 2011). On the contrary, ion fast ignition
(IFI) offers the advantages of a well known and well behaved ion-plasma
interaction, a much more localized energy deposition and a stiffer
ion transport with the possibility of beam focusing. A review of
the IFI current status can be found in Fern\'andez et al. (2014).

Recently, quasi-mononenergetic ion beams have been proposed for IFI
(Fern\'andez et al. 2009). These ions can be generated by either
laser-breakout afterburner (BOA) (Yin et al. 2007, Fern\'andez et al. 2009,
Yin et al. 2011a, Yin et al. 2011b, Hegelich et al. 2013,
Jung et al. 2013), radiation pressure acceleration (RPA)
(Macchi et al. 2005, Robinson et al. 2008) or ion solitary wave
acceleration (ISWA) (Yin et al. 2011c, Jung et al., 2011) schemes,
which use very thin foils, tens
or hundreds nanometers in thickness, illuminated by sub-picosecond
laser pulses. In addition, Weng et al. have proposed an {\it in-situ}
hole boring IFI scheme to generate quasi-monoenergetic ions in
overdense plasmas (Weng et al. 2014). Quasi-monoenergetic ions
have several advantages over ions with Maxwellian energy
distributions such as their better coupling with the compressed
fuel (Honrubia et al. 2009) and the possibility to place the
ion source far from the fuel without using re-entrant cones
(Fern\'andez et al. 2009). The progress of IFI driven by
quasi-monoenergetic ions has been summarized by Hegelich et al.
(2011), where it is pointed out the experimental
demonstration of the required i) particle energies (400-500 MeV),
ii) energy spreads (10-20\%) and iii) conversion efficiencies
($>$10\%). The chances over the forthcoming years to achieve 
experimentally all these simultaneously are high, very likely in new laser
facilities. The focusing of ion beams to small enough spots
($<$30 $\mu$m) still awaits experimental confirmation.

It is relevant here to recall the substantial progress on
proton focusing in cone targets reported recently
by Bartal et al. 2012. It is based on the setting up of an electron
sheath at the cone walls, that avoids the expansion of
co-moving electrons and, thus, contributes to proton beam
focusing. The spot size obtained in the experiments
carried out by Bartal et al. (2012) suggests that it may
be possible to focus protons into the spots required
for IFI. 

Here, we report on using heavier quasi-monoenergetic
ions and propose an improved target irradiation scheme in order
to minimize the ignition beam requirements and
the number of ions to be accelerated. This may be critical
for the successful demonstration of IFI in future
high-power laser facilities such as HiPER (Dunne 2006).

Full numerical modeling of the IFI scheme, from beam
generation to fuel ignition, is not possible with
the existing computer resources because it requires the
integration of physical processes with very different
spatial and temporal scales. Here, we use a simplified
model that assumes a rather ideal initial ion distribution
function and analyze the ion energy deposition and fuel
ignition in an ideal IFI scenario with a precompressed
fusion capsule.
 
This paper is organized as follows. After a short presentation
of the computational model, the ignition energies of collimated
beams of different ion species impinging on a precompressed
DT target are presented as a function of the ion energy.
Thus we can show what is the optimal ion species for different
ion acceleration conditions. Next, a new target irradiation
scheme with an array of focused beams is proposed. This scheme
reduces substantially the ignition energies, provided that
ion beams can be focused into 10 $\mu$m diameter spots. The
sensitivity of the ignition energies to the ion focal spot
size is analyzed. Finally, conclusions and future work are
briefly outlined.

\begin{figure}
\begin{center}
\includegraphics[width=.4\textwidth]{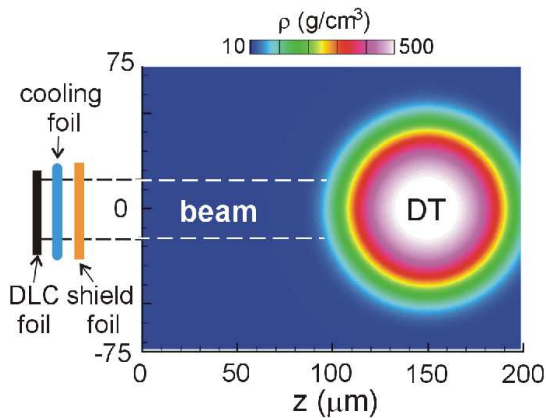}
\caption{\label{fig:1} Density map of the fuel configuration
considered in this work. The beam is perfectly collimated along
the distance $d=$ 5 mm from the ion source to the simulation box.}
\end{center}
\end{figure}

\section{Simulation model}

We assume cylindrical ion beams impinging on an ideal
configuration of compressed DT fuel with the
super-Gaussian density distribution
$\rho(r) = \rho_{peak} \exp[-\log2 (r/R)^4]$,
where $r$ is the distance to the center,
$\rho_{peak}=$ 500 g/cm$^3$, $R$= 82 $\mu$m and the
areal density is $\rho$R = 2 g/cm$^2$.
Two configurations of the ion beams are considered
in this work: perfectly collimated (constant flux
within its cross section) and focused ion beams, that
are studied in Sections 3 and 4, respectively
(Honrubia et al. 2013).
The simulation box used for perfectly collimated beams
is shown in Fig.~\ref{fig:1}. Ions come from the left
and propagate towards the dense core through a low
density plasma at 10 g/cm$^3$. In the particular
case shown in Fig.~\ref{fig:1} of quasi-monenenergetic
ions sited 5 mm far from the simulation box, a
re-entrant cone is not necessary and therefore it has
not been included in the calculations. Shield and
cooling foils (Huang et al. 2011) have not been included
either because they are too thin (around 10 $\mu$m
each) to induce relevant ion energy losses.
On the other hand, because the fuel is stagnated at the
time of maximum compression, we assume that the DT is
initially at rest with an uniform temperature of
100 eV. This corresponds to a ratio between the
plasma pressure and the Fermi pressure of 2.21 at
the peak density of 500 g/cm$^3$.

Calculations have been performed with the 2-D radiation-hydrodynamics
code SARA, that includes flux-limited electron conduction, multigroup
radiation transport, ion energy deposition, DT fusion reactions and
$\alpha$-particle transport (Honrubia, 1993a; 1993b). This
code is also coupled to a hybrid model for fast electron transport
in electron-driven fast ignition (Honrubia et al., 2006).

\begin{figure}
\begin{center}
\includegraphics[width=.35\textwidth]{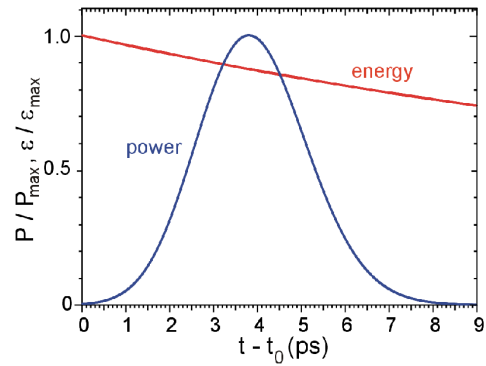}
\caption{\label{fig:2} Beam power and ion kinetic
energy at the left surface of the simulation box as
a function of time. The incoming beam has a 5 kJ energy
and a Gaussian spectrum centered at $\epsilon_0$ = 450 MeV
with an energy¡ spread of $\delta \epsilon/\epsilon_0$ =
0.1 (FWHM). The distance from the ion source to the left
surface of the simulation box is $d$ = 5 mm. Other
parameters shown in the plot are t$_0$ = 55 ps,
$\epsilon_{max}$ = 515 MeV and P$_{max}$ = 1.5 PW.
The beam power and ion energy on target have been
obtained by using the formulas shown by Temporal (2006).} 
\end{center}
\end{figure}

\subsection{Ion pulse on target}

We assume that ions are generated instantaneously with
a Gaussian energy distribution
$f(E) \propto \exp\{-\log2 [(\epsilon-\epsilon_0)/\delta \epsilon_0]^2\}$
where $\epsilon_0$ is the ion mean energy and
$\delta \epsilon / \epsilon_0$ is the energy spread, chosen as 0.1
in this work. We refer to this distribution as "quasi-monoenergetic"
throughout the paper. Instantaneous emission of the beam
ions is assumed because the time of flight spread ($\approx$ 3 ps)
is much longer than the ion acceleration times found for
the BOA scheme. Yin et al. (2011a, 2011b)
have shown that ions are accelerated mainly between 
the time of the foil relativistic transparency $n_e \approx \gamma n_c$
and the time when the target becomes undercritical ($n_{e,peak} \approx n_c$),
where $n_e$ and $n_c$ are the electron density and the critical density,
respectively, and $\gamma$ the relativistic Lorentz factor. For 
a 100 nm thick DLC (diamond like carbon) foil illuminated by a peak
laser irradiance of $5.2\times10^{20}$ W/cm$^2$, Yin et al.
obtained an enhanced acceleration time of 400 fs, still negligible when
compared with the time of flight spread shown in Fig.~\ref{fig:2}.
Anyhow, simulations taking into account the pulse duration mentioned
above show no differences when compared with the instantaneous
emission used here.

Beam power and ion kinetic energy on target of a 5 kJ
quasi-monoenergetic carbon ion beam generated at a
distance $d$ = 5 mm are depicted in Fig.~\ref{fig:2}.
This distance has been chosen in order to have a peak
power about 1.5 PW and a pulse duration on target of
$\approx$ 3 ps (FWHM). Note that the small time spread
of quasi-monoenergetic ions allows to place the source
at much higher distances than Maxwellian ions. Because
beam focalization over such distances may be difficult,
a number of techniques have been proposed for that.
Some of them are: i) ballistic transport (Key 2007,
Patel et al. 2008), ii) focusing by fields self-generated
in hollow microcylinders by intense sub-picosecond laser
pulses (Toncian et al., 2006) and iii) focusing by magnetic
lenses (Schollmeier et al. 2008, Harres et al. 2010,
Hofmann et al. 2011).

Focusing of ions with Maxwellian spectra has been
studied over the last years. Kar et al. (2008)
have demonstrated experimentally proton beam focusing
by using foil rectangular or cylindrical hollow lens
attached to a foil target. Offermann et al. (2011)
have shown theoretically and experimentally that ion
divergence depends on the thermal expansion of
the co-moving hot electrons, resulting in
a hyperbolic ion beam envelope. Using these results,
Bartal et al. have shown experimentally an enhanced
focusing of TNSA-protons in cone targets, inferring
spot diameters about 20 $\mu$m for IFI conditions,
well under the 40 $\mu$m spots required (Bartal et al. 2012).
However, as it has been shown recently
by implicit PIC simulations (Qiao et al. 2013), the cone wall
focusing mechanism may reduce substantially the laser-to-proton
conversion efficiency and, thus, the ion energy for the long
pulses required in IFI. This can be mitigated by special
designs of cone walls including insulator materials in
order to reduce the electron flow between target and cone
(Qiao et al. 2013).

For quasi-monoenergetic ions and, in particular, for those
accelerated by the BOA scheme, Huang et al. (2011) have proposed
recently the use of a second foil (see Fig.~\ref{fig:1})
to cool down the co-moving electrons and to absorb the
trailing laser pulse that pass through the main foil after
it becomes relativistically transparent (Huang et al. 2011a and 2011b).
The electron cooling obtained in this way reduces substantially
both the beam divergence and the energy spread, and can be used
together with the methods mentioned above for beam focusing
in order to get the required spot size on target.
In addition, heavier ions are less sensitive than protons
to the co-moving electron expansion.

\subsection{Ion stopping}

\begin{figure}
\begin{center}
\includegraphics[width=.4\textwidth]{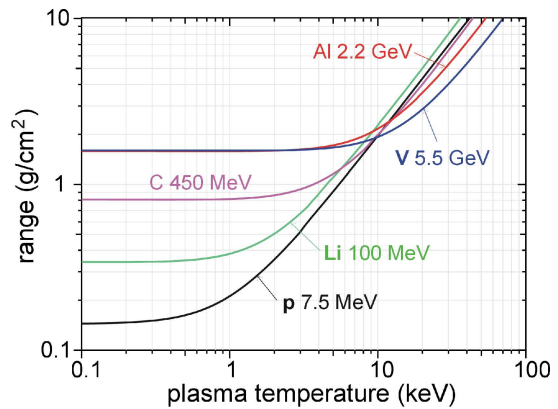}
\caption{\label{fig:3} Range of monoenergetic ions
in DT at 300 g/cm$^3$ as a function
of the plasma temperature.}
\end{center}
\end{figure}

Ranges of different ion species with typical energies for FI
are shown in Fig.~\ref{fig:3} as a function of the DT plasma temperature.
We have used the standard stopping power formalism for classical plasmas
(Trubnikov 1965, Honrubia 1993b) which predicts range lengthening when plasma
electron thermal velocities are comparable to fast ion velocity.
The range lengthening effect is important for ions with Maxwellian energy
distributions placed far from the target. It balances the
reduction over time of the ion energy incident on the fuel,
keeping the ion range almost constant (Temporal et al. 2002).
On the contrary, as the ion energy on
target is approximately constant for quasi-mononenergetic ions,
range lengthening increases the volume heated by the beam and
thus the ignition energies. Fortunately, the importance of range
lengthening is lower for heavier ions, as shown in Fig.~\ref{fig:3}.
For instance, protons increase their range by one order of magnitude
when the DT temperature rises from 0.1 to 10 keV, while heavier ions
such as aluminum or vanadium have an almost constant range for those
temperatures. Thus, a better coupling of heavier
ion beams with the plasma should be expected (Honrubia et al. 2009).
Note that the range of the ions shown in Fig.~\ref{fig:3} at 10 keV
is about 2 g/cm$^2$, higher than the 1.2 g/cm$^2$ requested for FI.
However, as the ions deposit all their energy before such high
temperatures are reached, the deposition range is substantially
shorter in the cases analyzed here, as shown in Section 3.1.

It is worthwhile pointing out again that because the heavier the
ion, the higher the energy they carry for a given range,
the number of ions required for ignition decreases with the atomic
number. It is lower by orders of magnitude for
vanadium ions than for protons.

\section{Ignition energies for perfectly collimated beams}

\subsection{Energy deposition}

\begin{figure}
\begin{center}
\includegraphics[width=.35\textwidth]{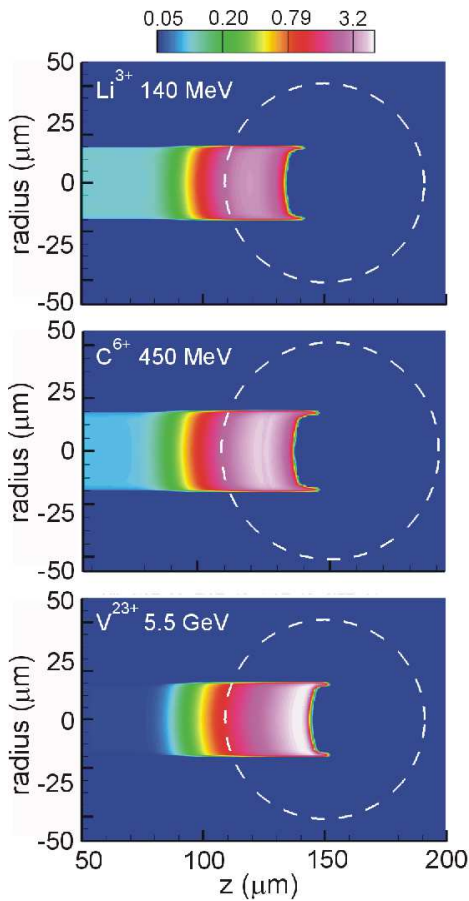}
\caption{\label{fig:4} Energy density in units of
10$^{11}$ J/cm$^3$ deposited in the target of Fig.~\ref{fig:1} by
8.5 kJ quasi-monoenergetic beams (a) 140 MeV lithium ions,
(b) 450 MeV carbon ions and (c) 5.5 GeV vanadium ions.
The distance $d$ from the ion source to the left surface of the
simulation box is 5 mm for all beams. The dashed curves show
the initial position of the density contour $\rho=$ 250 g/cm$^3$.}
\end{center}
\end{figure}

The energy deposition by lithium, carbon and vanadium ion beams
with different kinetic energies are compared in Fig.~\ref{fig:4}.
These energies have been chosen as those which minimize the
ignition energies, as discussed in Section 3.3. Energy deposition
has been computed by assuming that ions propagate in a straight line
and thus neglecting the angular scattering by plasma ions, that is
important for low energy ions only.
Since the pulse duration is short enough compared to the resulting
DT expansion (with exception of the cylindrical beam edge, where a shock
wave is launched into the cold plasma), the volume heated
by the beam is determined mostly by the ion energy and the range
lengthening effect. The three beams shown in Fig.~\ref{fig:4}
have a similar penetration into the dense core, almost up to
its center. It corresponds to an areal density of around
1.6 g/cm$^2$, higher than Atzeni's prescription of
1.2 g/cm$^2$ (Atzeni 1999, Atzeni et al. 2002) due to the
deposition in the coronal plasma.
Lithium ions show a much less localized energy deposition
than heavier ions, such as vanadium, due to their higher
range lengthening. Note also that the energy deposition
in the coronal plasma decreases for heavier ions. Thus,
the coupling efficiency (defined as the energy deposited
in the plasma at densities higher than 200 g/cm$^3$)
should be higher for heavier ions. In particular, for
the lithium, carbon and vanadium beams of Fig.~\ref{fig:4},
the coupling efficiencies are 0.74, 0.81 and 0.89, respectively,
well over the coupling efficiencies found for ions with a
Maxwellian energy distribution (Honrubia et al., 2009).

\subsection{Optimal beam radius}

\begin{figure}
\begin{center}
\includegraphics[width=.45\textwidth]{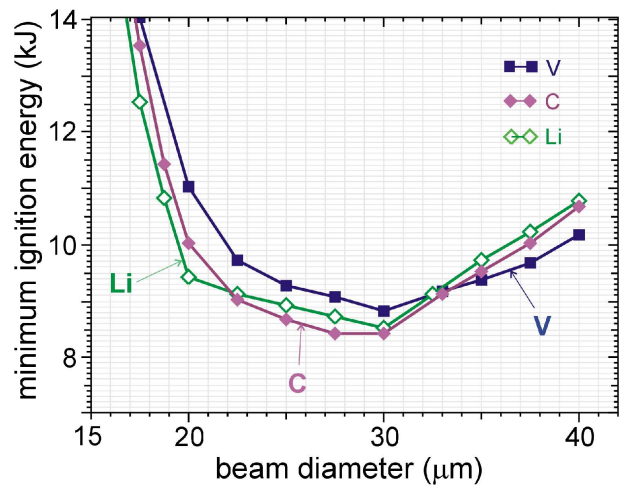}
\caption{\label{fig:5} Minimum ignition energies
of the target shown in Fig.~\ref{fig:1} heated by
100 MeV lithium, 450 MeV carbon and 5.5 GeV vanadium
ions as a function of the beam diameter. All the
beams have an energy spread of 10\%. The
source-core distance $d$ is 5 mm.}
\end{center}
\end{figure}

Ignition energies $E_{ig}$ as a function
of the beam diameter for different ion beams are
shown in Fig.\ref{fig:5}. They have been
obtained as the minimum beam energy for which the
thermonuclear fusion power has an exponential or higher
growth sustained in time. The ignition
energies of all beams are quite close and have a similar
variation with the beam diameter. The lowest ignition energy,
8.3 kJ, is obtained for 450 MeV carbon ions. In all cases,
the ignition energies rise for lower and higher
beam diameters, showing almost a plateau for diameters
between 20 and 40 $\mu$m, which determines
the focusing requirements for IFI. For beam diameters
lower than 20 $\mu$m, the energy density deposited is very
high, leading to a strong plasma expansion and range
lengthening with the subsequent increase of ion penetration.
Ions can even pass through the compressed fuel and escape
off the rear surface. For beam diameters higher than
40 $\mu$m, the ignition energies $E_{ig}$ rise again due
to the larger spot volume that has to be heated.
However, the increase of the ignition energies is
less than proportional to that volume due to the
reduction of $\alpha$-particle losses from large spots.

\subsection{Optimal ion energies}
 
\begin{figure}
\begin{center}
\includegraphics[width=.46\textwidth]{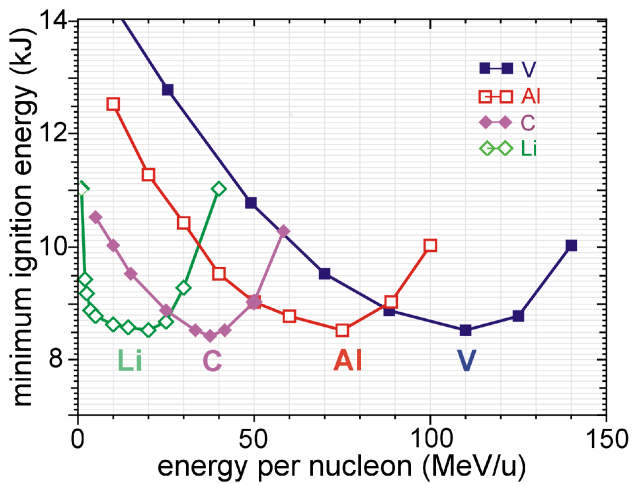}
\caption{\label{fig:6} Minimum ignition energies
of the target shown in Fig.~\ref{fig:1} heated by
quasi-mononenergetic lithium, carbon, aluminum
and vanadium ions as a function of the mean kinetic energy
per nucleon. The source-target distance $d$ is 5 mm, 
the energy spread $\delta \epsilon/\epsilon$ = 10\%
and the beam diameter 30 $\mu$m in all cases.}
\end{center}
\end{figure}

The ignition energies for different beams as a function of
the ion energy per nucleon are shown in Fig.~\ref{fig:6}.
It is interesting to point out the remarkable result that all
the ions have similar ignition energies, around 8.5 kJ for
the optimal kinetic energies $\epsilon_0$. These optimal energies
increase with the atomic number, being 140 MeV for lithium,
450 MeV for carbon, 2 GeV for aluminum and 5.5 GeV for
vanadium ions, scaling, approximately, as $Z^2$. This
scaling is important because it prescribes the optimal
ion type as a function of the available ion-accelerating electric
potential. The increase of the optimal ion kinetic energies with Z
was also reported by Albright et al. (2008) and 
Fern\'andez et al. (2008) within the context of IFI
capsule designs for NIF. 

The shape of the curves shown in Fig.~\ref{fig:6} can be explained
by taking into account that for low ion energies $\epsilon_0$
the pulse on target has a relatively low power, $P \propto \epsilon_0^{1/2}$
and a longer duration, $t_{pulse}\propto \epsilon_0^{-1/2}$ (Honrubia et al. 2009).
In this case, the pulse departs from the optimal one and the ignition
energies $E_{ig}$ increase. For high ion energies, $E_{ig}$
increases again due to the higher fuel mass heated by the ion beam.

As was mentioned in Section 2.2, the rise of the optimal
ion energy with the atomic number of fast ions leads to a
strong reduction of the number of ions required for ignition.
For instance, around 10$^{13}$ vanadium ions of 5.5GeV are
required to ignite the target shown in Fig.~\ref{fig:1},
which is three orders of magnitude lower than that required
for the standard proton fast ignition scheme (Roth et al. 2001).

\section{Ignition energies for arrayed beams}

\begin{figure}
\begin{center}
\includegraphics[width=.45\textwidth]{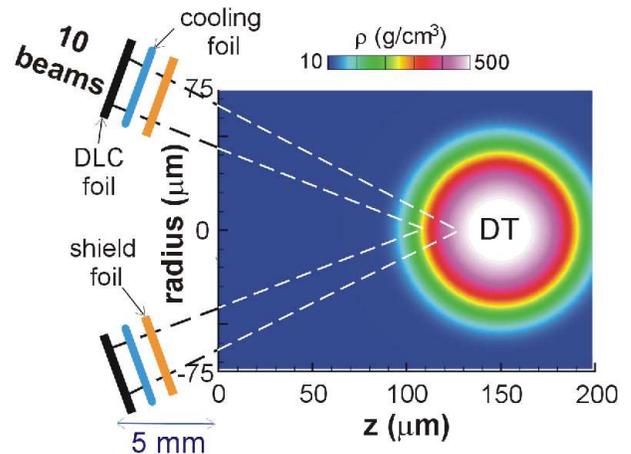}
\caption{\label{fig:7}
Sketch of the arrayed beams illumination scheme.
Ten laser beams impinge on a ring located 5 mm far from the 
dense core. The ion beams generated at the rear target surface
are focused on a 10 $\mu$m spot placed in the density ramp
in order to generate a hollow cone energy deposition pattern.}
\end{center}
\end{figure}

In order to reduce the IFI beam requirements,
we propose a new irradiation scheme aimed at
improving the beam-target coupling and igniting the
compressed core more efficiently than the single beam
scheme discussed in Section 3. In addition, it may
lead to a further reduction of the ion number.
Basically, the scheme consists of using a set of
beams generated far from the imploded core and 
focusing them into a spot placed in the density
ramp surrounding the core. Crossing beams generate
a hollow cone energy deposition pattern. The scheme is
shown in Fig.~\ref{fig:7}, where the ions are generated in a ring
placed 5 mm far from the core with a mean diameter of 2.7 mm and
a tilting angle of 15$^{\circ}$. In the reference case discussed
below, we consider $N=$ 10 ion beams generated in spots placed
symmetrically at the ring with a radius $r_{spot}=$ 20.3 $\mu$m
and area $S=\pi r_{spot}^2$ each. We assume that N = 10 beams
are sufficient to have an almost homogeneous ion ring on target,
as shown by Temporal et al. (2009). To illustrate the new
scheme and show the advantages of more tightly focused ion
beams, laser beam
parameters close to those used in recent BOA experiments and
simulations have been chosen. Thus, we assume a laser intensity
of $I_L=$ 10$^{21}$ W/cm$^2$ and a laser-to-ion conversion
efficiency $\eta=$ 10\%. The laser power is $P_L=N I_LS$,
the laser pulse energy is $E_L=E_{ig}/\eta$ and the laser pulse
duration is $\tau=E_{ig}/(N I_LS\eta)$.
For $E_{ig}$ = 6.5 kJ, one obtains $P_L (PW)$ = 13$\times N$,
$E_L$ = 65 kJ and $\tau (ps)$ = 5/$N$. For the reference case
$N=$ 10 ion beams, each has a power $P_{beam}=$ 1.3 PW and an energy
$E_{beam}=$ 0.65 kJ. The pulse duration would be $\tau=$ 0.5 ps,
 and the carbon ion mean energy 500 MeV. Lower beam powers can be
obtained with longer laser pulses, that have not been studied
theoretically nor experimentally yet.

\begin{figure}
\begin{center}
\includegraphics[width=.48\textwidth]{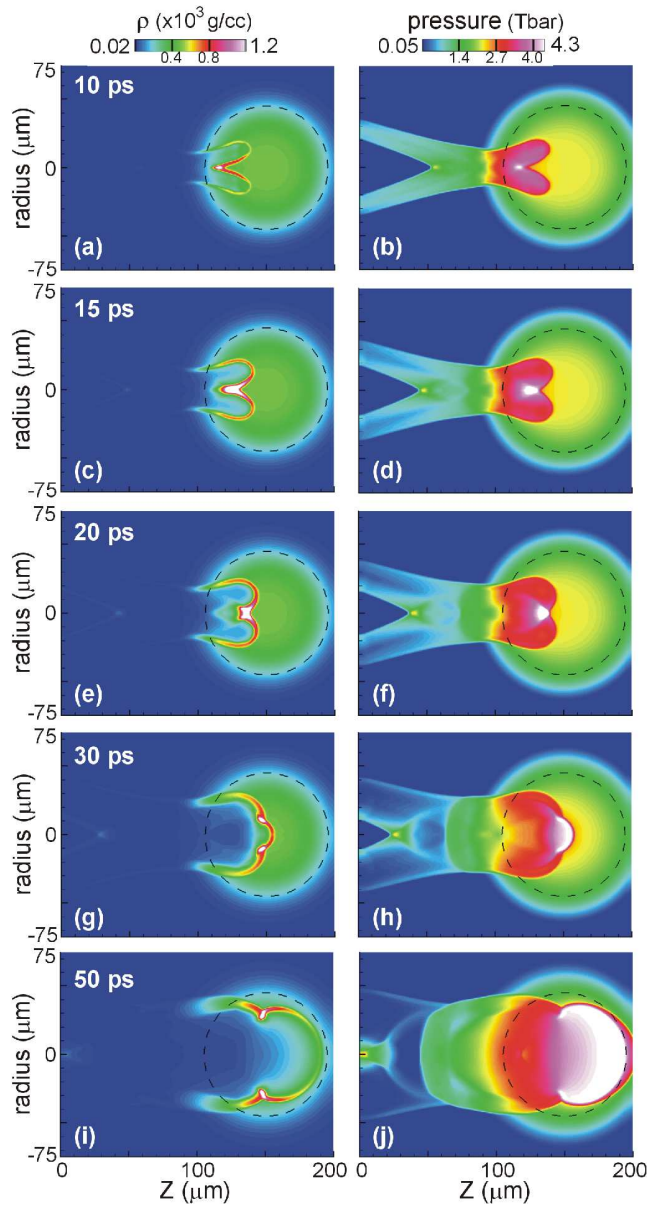}
\caption{\label{fig:8} Ion density (left panels) and pressure
(right panels) evolution of the target shown in
Fig.~\ref{fig:7} heated by carbon ion beams of 500 MeV
and $\delta \epsilon / \epsilon=$ 0.1 focused onto a 10 $\mu$m spot
sited at z=100 $\mu$m. Ignition is produced by the shock waves
launched by the ion beams that collide on the axis.}
\end{center}
\end{figure}

\begin{figure*}
\begin{center}
\includegraphics[width=.9\textwidth]{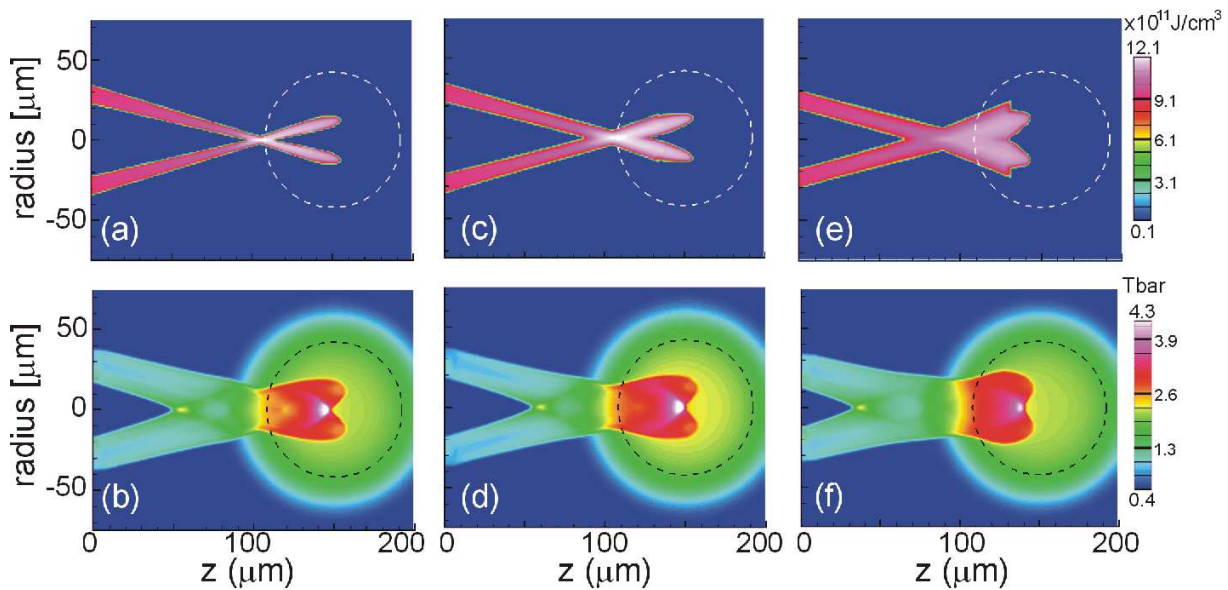}
\caption{\label{fig:9} Sensitivity of the target heating on
beam focusing. Energy deposition of 500 MeV carbon ion beam
is shown in the upper panels and DT pressure at 15 ps in the
lower panels. The focusing parameter $d_{foc}$, defined as
the overlapped minimum beam diameter on axis, and the ignition
energies $E_{ig}$ are: (a,b) $d_{foc}= 5 \mu$m, $E_{ig}=4.5$ kJ;
(c,d) $d_{foc}= 10 \mu$m, $E_{ig}=5.7$ kJ; and (e,f)
$d_{foc}= 15 \mu$m, $E_{ig}=7$ kJ.}
\end{center}
\end{figure*}

The evolution of the target shown in Fig.~\ref{fig:7} can be
summarized as follows. The beams are focused into a spot placed
in the density ramp such that the ions still have enough
energy to penetrate further into the compressed fuel [see
Figs.~\ref{fig:8}(a) and (b)]. Most
of the energy deposition in the high density fuel takes
place in a hollow cone. The dense fuel within the
deposition zone expands and launches a strong shock wave that
propagates towards the axis, where the shocks collide
and compress the fuel further to very high densities
($\approx$ 1200 g/cm$^3$) [see Fig.~\ref{fig:8}(c)].
The pressure peaks there [see Fig.~\ref{fig:8}(d)] and
increases its strength while propagating towards the right
[see Figs.~\ref{fig:8}(e) and (f)]. Ignition starts in this
high-pressure region [see Figs.~\ref{fig:8}(g) and (h)]
and propagates through the dense and cold fuel [see
Figs.~\ref{fig:8}(i) and (j)]. In this irradiation scheme,
ignition is produced by the collision of two shocks on the axis,
not by the direct heating of the fuel, being in this sense
a variant of the shock ignition scheme (Betti 2006). The
main advantage of this irradiation scheme is that ignition
can be achieved with substantially lower beam energies.
For instance, a beam of 500 MeV carbon ions focused onto
a 10 $\mu$m spot diameter at the depth of $z$ = 100 $\mu$m
on the axis requires 5.7 kJ for ignition, which is
approximately 2/3 of the energy required for a single beam,
as shown in Fig.~\ref{fig:5}. It is worthwhile to emphasize
that this improvement is not due to a sharper ion-beam
focusing {\it per se}, as clearly shown in Fig.~\ref{fig:5}
for a single beam. Contrary to the single-beam case,
the success of the array scheme depends on a sharper
focusing of the ion beams able to generate strong
converging shock waves. For instance, if the beams could be
focused into a 5 $\mu$m spot, the ignition energy would be
reduced to 4.5 kJ, roughly a half of that obtained
for a single beam. The dependence
of the ignition energy on the spot diameter is shown in
Fig.~\ref{fig:9}. For diameters as large as 15 $\mu$m, the
focusing beams scheme still reduces the ignition energy
by $\approx$20\% when compared with the single beam scheme.
For larger diameters, the differences between both schemes are
not high enough to balance the difficulties of beam focusing.
Thus, its practical implementation is limited by the possibility of
beam focusing into 10 $\mu$m spots, smaller than those required
for single beams (30 $\mu$m), from distances of millimeters,
which is a challenging task.

The focused beams scheme presented above has some advantages when
compared with that described in Temporal et al. (2008) and (2009)
for proton FI. In this latter scheme, the imploded target is first irradiated
by a number of proton beamlets with an annular set up and a total energy of
1 kJ followed, after a time delay, by a second cylindrical beam of 7 kJ.
The penetration of the central part of the cylindrical beam is blocked by
the higher densities generated on the axis by the annular beamlets.
Meanwhile the outer part of the cylindrical beam penetrates further and
generates a cylindrical shock that collides on the axis and ignites the
DT fuel ahead the energy deposition of the central part of the ion beam.
On the contrary, in our scheme,
the shocks are produced directly by the whole beam energy deposition
without any blocking effect and thus the scheme should be more
effective. In addition, in the scheme of Temporal et al. (2009),
ignition is very sensitive to the time delay between the annular
beams and the main cylindrical beam while in ours all beams are
fired simultaneously. It is also worth noting that the focusing
requirements are similar in both schemes because the diameter
of each beamlet is 10 $\mu$m, just the same than the spot
required for our beams array.

\section{Conclusions}

In order to extend the possibilities of IFI with quasi-monoenergetic
ions, we consider different ion species and propose a new irradiation
scheme with arrayed beams. Specifically, fast ignition by quasi-monoenergetic
lithium, carbon, aluminum and vanadium ions have been analyzed for a
simplified DT fuel configuration with a peak density of
500 g/cm$^3$. Using middle or heavy ion beams has
the advantage of reducing substantially, by orders of magnitude,
the number of ions required for ignition. Ideal ion beams with
a uniform flux within its cross section and an energy spread
of 10$\%$ perfectly focused into such a configuration have been analyzed.
Simulations show that the minimum ignition energies, about 8.5 kJ,
are similar for the ions studied here despite being obtained with
very different kinetic energies. We should point out
that those ignition energies have to be considered as a lower
limit due to the strong assumptions made. Anyhow, taking into
account laser-to-ion conversion efficiencies around 10\%
or higher found so far (Hegelich et al. 2011), it would be possible
to fast ignite a pre-compressed target with laser energies of about
100 kJ. It is worth mentioning the high laser-to-proton conversion
efficiencies (15\%) measured recently in target normal sheath
acceleration experiments (Brenner et al. 2014).

In addition to the collimated beam studies,
we propose a new target irradiation scheme based on an array of focused
quasi-monoenergetic ion beams. As a single 100kJ short-pulse
laser cannot easily envisaged, the single beam scheme would be
based on the combination/superposition of multiple beamlets to get
the required energy, in a similar way that the ion beam array
proposed here. The difference is that in our scheme the beamlets
should be focused onto a spot sited on the target density ramp.
That may be a better way of deploying individual FI laser beams
(or individual clusters of laser beams).

Our results show that the arrayed beam
scheme reduces substantially the ignition energies obtained
for single beams, provided that the ion beams can be
adequately focused into 10 $\mu$m spots.
Of course, to achieve such focusing will require a substantial
reduction of the transverse temperature of  co-moving electrons,
for instance, by using a second cooling foil (Huang et al. 2011),
which shall be used together with one of the focusing techniques
outlined in Section 2.1. The price to be paid for
the reduction of the ignition energy would be the 3D focusing of
the beam array, which may have a similar or even higher difficulty
than using lasers with energies above 100kJ.

Future studies of IFI will include realistic beam and target
configurations, and longer laser pulses in order to obtain
more accurate estimations of the laser and ion beam requirements.

\begin{acknowledgments}
This work was partially supported by the research grant ENE2009-1168
of the Spanish Ministry of Education and Research, was undertaken as
part of the HiPER preparatory project, and used resources and technical
assistance from the CeSViMa HPC Center of the Polytechnic University
of Madrid. This work was also partially supported by the LDRD
program at Los Alamos National Laboratory, and by the US DOE.
\end{acknowledgments}

\def\bibindent{1em}

\end{document}